\documentclass[structabstract]{aa}  
\usepackage[authoryear]{natbib}
\usepackage{graphicx}
\usepackage{txfonts}

\def\smallsun{\hbox{$_\odot$}}
\def\degr{\hbox{$^\circ$}} 

\def\arcsec{\hbox{$^{\prime\prime}$}}
\def\cm3{cm$^{-3}$}
\def\ebv{$E_{\mathrm{B-V}}=$}

\begin{document} 


\title{Dust properties in the galactic bulge \thanks{Based on observations with
the {\em Spitzer} Space Telescope, which is operated
     by the Jet Propulsion Laboratory, California Institute of Technology.}}

\author{S.R.\,Pottasch\inst{1}, \and J.\,Bernard-Salas\inst{2}}

\offprints{pottasch@astro.rug.nl} 

\institute{Kapteyn Astronomical Institute, P.O. Box 800, NL 9700 AV
Groningen, the Netherlands  \and Institut d’Astrophysique Spatiale, Paris-Sud 
11, 91405 Orsay, France}

\date{Received date /Accepted date}

\abstract
{It has been suggested that the ratio of total-to-selective extinction R$_V$ in
dust in the interstellar medium differs in the Galactic bulge from its value in
the local neighborhood.}
{We attempt to test this suggestion.}
{The mid-infrared hydrogen lines in 16 Galactic bulge PNe measured by the 
Spitzer Space Telescope are used to determine the extinction corrected 
H$\beta$ flux. This is compared to the observed H$\beta$ flux to obtain the 
total extinction at H$\beta$. The selective extinction is obtained from the
observed Balmer decrement in these nebulae. The value of R$_V$ can then be
found.}
{The ratio of total-to-selective extinction in the Galactic bulge is 
consistent with the value R$_V$=3.1, which is the same as has been found in 
the local neighborhood.}
{The suggestion that R$_V$ is different in the Galactic bulge is incorrect. 
The reasons for this are discussed.}

\keywords{ISM: abundances -- planetary nebulae: individual:
  \object{He2-367 \& He2-250 \& H2-11 \& He2-262 \& H1-15 \& M1-38 \& M1-37 \&
H2-20 \& M2-31 \& He2-260 \& M2-5 \& M2-10 \& M1-27 \& H1-43 \& H1-20 \& M3-44}
 -- Infrared: ISM: 
lines and bands}

\authorrunning{Pottasch et al.}
\titlerunning{Dust properties....}  

\maketitle

\section{Introduction}
 
Interstellar extinction is a well studied but not completely understood 
phenomen. Beside studying the properties of the particles which cause the
extinction, it is necessary to know the correction for extinction as a 
function of wavelength since individual measurements must be corrected for the
effects of extinction.This has led to several Galactic extinction curves; the 
most commonly used are those of Savage \& Mathis\,\cite{sm}, Seaton\,\cite{seaton} and Cardelli et al.\,\cite{ccm}. These curves extend over the complete 
wavelength range through the ultraviolet to about 1000\AA. All three curves 
cited are quite similar and are generally used as an average Galactic 
extinction curve to correct observed spectra for the effects of extinction. 
While uncertainties exist in these curves in the ultraviolet part of the 
spectrum, the uncertainties in the visible and infrared regions are small, 
almost certainly less than 5\% (eg Fitzpatrick \cite{fitz1}). An additional
extinction curve has been proposed for the near and mid-infrared spectral region
by Chiar \& Tielens\,\cite{chiar}.

In the specific case of correcting the spectra of planetary nebulae for the
effects of extinction two methods are used, both of which are based on the 
average Galactic extinction curve. The first makes use of a comparison of the 
observed Balmer decrement with that expected from optically thin nebulae using
the theoretical values given by Hummer \& Storey\,\cite{hummer}. Ususally 
fitting the H$\alpha$/H$\beta$ ratio is given the most weight. The best fit 
between the observed and theoretical Balmer decrement is usually given by the 
parameter C$_{bd}$ which is the log of the value by which the observed H$\beta$
must be increased to correct it for interstellar extinction. C$_{bd}$ is a weak
function of T$_{e}$, the electron temperature, and R$_{V}$, the ratio of total-
to-selective extinction at visual wavelengths. A value of R$_{V}$=3.1 is usually used. In principle C$_{bd}$ is also a function of the electron density, but it 
is such a weak function for the densities found in planetary nebulae that we 
will ignore it here.

The second method for correcting for the effects of extinction is to use the 
observed radio continuum emission. The radio emission is dependent on the 
product of the density of ionized hydrogen and the electron density in the same
way as the H$\beta$ emission is, so that the ratio is essentially density 
independent. The value of unreddened H$\beta$ thus obtained may be used to 
compute C$_{rad}$, similar to the value of C computed above but now derived from
the radio continuum flux density. Note that C$_{rad}$ is a weak function of 
T$_{e}$. It also depends on the abundance of (doubly) ionized helium which 
contributes to the electron density. But C$_{rad}$ is independent of R$_{V}$. 
Note that it is also assumed that the nebula is optically thin to the radio 
emission so that all emission produced is measured. Because the geometry of the
nebula is not known, the optical depth cannot be accurately computed. Because 
it is inversely proportional to the square of the frequency, the radio 
measurements should be made at the highest possible frequency.
 
Since C$_{bd}$ and C$_{rad}$ are by definition equal and only C$_{bd}$ depends on 
R$_{V}$, information concerning this quantity can now be found if the Balmer 
decrement, the radio continuum flux density, the electron temperature and the 
amount of helium and its ionization are known. This method has already been 
applied to PNe in the Galactic bulge 20 years ago by Tylenda et al.\,\cite{tylenda} and Stasinska et al.\,\cite{stasinska}. In the 
first of these articles the authors found that there was fair agreement between
C$_{bd}$ and C$_{rad}$ for small values of extinction, but for large extinctions 
C$_{rad}$ is smaller than C$_{bd}$. In the first paper they explain the difference as due to underestimates of the radio emission. In the second article 
they express more confidence in the radio measurements and now explain the 
difference as being due to a lower value of R$_{V}$ for the more distant 
nebulae. While these authors do not specifically refer to the Galactic bulge, 
about 30\% to 40\% of the PNe they use are Galactic bulge objects. Other 
authors (eg Hultzsch et al.\,\cite{hultz} and Ruffle et al.\,\cite{ruffle}) 
use this as evidence for a possibly lower value for R$_{V}$ in the Galactic 
bulge.

The calculation of C$_{bd}$ and C$_{rad}$ is uncertain as these authors admit. 
First of all the electron temperature of the individual nebulae is unknown, so 
that a value of 10000 K was always used. Secondly the helium abundance is 
unknown so that a 10\% helium abundance is assumed, all in singly ionized form.
Thirdly, the measurements of the Balmer decrement are sometimes of uncertain 
quality; in the few cases where multiple observations of a single object have
been made conflicting results are sometimes obtained. Finally no correction 
has been attempted for possible optical depth effects in the 6\,cm radio 
continuum except that objects with possibly high brightness temperature were 
removed from the sample. This method is not very reliable, as the above 
authors remark, because the surface brightness of PNe 'is usually far from 
uniform and the estimate of the brightness temperature is very sensitive to 
the adopted nebular diameter'. The sample used by Stasinska et al.\,\cite{stasinska} contained somewhat less than 199 Galactic bulge PNe, while the sample 
used by Ruffle et al.\,\cite{ruffle} contained 12
Galactic bulge PNe. This latter sample, which reached the same conclusion as 
that of Stasinska et al.\,\cite{stasinska}, contains the same uncertain 
approximations.

There is a better method for comparing the Balmer decrement extinction with 
that determined from the total H$\beta$ extinction. In this method the hydrogen
lines in the mid-infrared are measured; in this part of the spectrum the 
interstellar extinction is very small. The intrinsic H$\beta$ intensity is then
derived from the mid-infrared line intensiy by the theoretical relations of
Hummer \& Storey\,\cite{hummer}. To measure the mid-infrared lines we use the
spectrograph on board the {\em Spitzer} Space Telescope which measures several
hydrogen lines, the strongest being the H(6-5) transition at 7.46 $\mu$m
(Paschen $\alpha$) and, with a higher spectral resolution, the H(7-6) transition at 12.37 $\mu$m. The log of this predicted value of H$\beta$ 
intensity divided by the observed H$\beta$ intensity is called C$_{ir}$ and may 
be directly compared with C$_{bd}$ to determine R$_{V}$. In this way the radio
continuum measurements are replaced by the infrared hydrogen line measurements;
the radio measurements are no longer necessary to determine R$_{V}$.

In section 2 the infrared observations are presented and discussed, resulting 
in the determination of C$_{ir}$. In section 3 the Balmer decrement is discussed
resulting in a value of C$_{bd}$. A comparison of these two values is made in
section 4, which leads to a conclusion about R$_{V}$. At the same time C$_{rad}$ 
is computed and discussed.

\section{The infrared spectrum}

\subsection{Observations}

We have selected 16 planetary nebulae for measurement, which are almost 
certainly in the Galactic bulge. They are all within 10\degr~of the galactic 
center, they have low intrinsic radio continuum flux density and all but one
have very high velocity with respect to the sun. The observations were made 
using the Infrared Spectrograph (IRS, Houck et al. \cite{houck}) on board the 
{\em Spitzer} Space Telescope with AOR keys both on target and on the 
background. The reduction of the spectra reported by Gutenkunst et al.\,\cite{guten} (see listing below) started from the {\em droop} images and is idscussed 
in detail in that paper. The results given by Gutenkunst et al.\,\cite{guten}
are shown in Table 1. For the remaining extractions {\em droop} images are
replaced by optimal extractions. The details of this reduction procedure are 
given in Pottasch \& Bernard-Salas (in prep.). 

Two of the three diaphragms used have high resolution: the short high module 
(SH) measures from 9.9$\mu$m to 19.6$\mu$m and the long high module (LH) from 
18.7$\mu$m to 37.2$\mu$m. The SH has a diaphragm size of 4.7\arcsec x 11.3\arcsec, while the LH is 11.1\arcsec x 22.3\arcsec. The diameters of the PNe being
measured are shown in Table 1. These are taken from a variety of sources, the 
most reliable are those of Sahai et al.\cite{sahai}. For the other PNe the VLA
radio continuum half width measurements listed by Acker et al.\cite{acker} are 
given. The diameters are always considerably smaller than the LH
diaphragm but may not always be smaller than the SH diaphragm. To correct for
possible missing intensity in the SH diaphragm we examine the continuum 
intensity of LH and SH in the region of wavelength overlap at 19$\mu$m. These 
continua should be equal if the entire nebula is being measured in the SH
diaphragm. If this is not so all intensities in the SH diaphragm are increased 
by a factor which make the continua equal. The corrections are small for these
nebulae, ususally between 1.1 and 1.3.

The third diaphragm is a long slit which is 4\arcsec~wide and extends over the 
entire nebula. This SL module measures in low resolution between 5.5$\mu$m and 
14$\mu$m. These spectra are normalized by making the strong lines in common 
between the SL and SH modules agree. Especially important is the agreement of 
the \ion{[S}{iv]} line at 10.51$\mu$m and the \ion{[Ne}{ii]} line at 12.82$\mu$m. Because all of the PNe measured are small, the corrections are likewise small.
The resultant spectra may now be plotted. We have not done this here because
good plots are already in the literature. Plots of He2-250, H2-11, He2-262, 
H1-15, H2-20, M2-31, M2-5, M2-10 and H1-20 may be found in Gutenkunst et al.\cite{guten}; plots of He2-260, M1-27, H1-43 and M3-44 can be found in Perea-Calderon et al.\cite{perea}; plots of M1-38 and M1-37 are found in Stanghellini et al.
\cite{stang}. The excellent quality of the emission lines is apparent on the 
spectra.
 
The intensities of the 12.37$\mu$m line (SH diaphragm) and the 7.46$\mu$m line 
(SL diaphragm) are then be measured using the gaussian line-fitting routine. 
The resultant values are shown in cols.3 and 4 of Table 1. The 3$\sigma$ errors
of measurement are always less than 20\%. The 12.37$\mu$m line has the more 
accurate intensity of the two lines in spite of the fact that it is intrinsically the weaker line. There are several reasons for this. First of all the  
12.37$\mu$m line is measured with a much higher resolution (about 600) compared
to the 7.46$\mu$m line (about 90). Secondly the measurement of the 7.46$\mu$m 
line is difficult because it falls in the middle of overlapping orders. Thirdly
the 12.37$\mu$m line is often measured both at high resolution (SH) and at low 
resoluton (SL) and very similar intensities are found. For these reasons we 
give the value of the intrinsic H$\beta$ found from the 12.37$\mu$m line
considerably more weight.  

The H$\beta$ flux is found from the infrared hydrogen lines using the 
theoretical ratios of Hummer \& Storey\,\cite{hummer}. It must be remembered 
that the 12.37$\mu$m line is a blend of the H(7-6) transition with the H(11-8) 
line, while the 7.46$\mu$m line is a blend of H(6-5) with H(8-6) and H(17-8).
These blend are subtracted when deriving H$\beta$. Futhermore the ratios of the
various hydrogen lines are weakly dependent on the electron temperature. The 
values of T$_{e}$ used are given in col.5 of Table 1 and are taken from 
Pottasch \& Bernard-Salas (in prep.). The values of H$\beta$ thus found
are given in cols.6 \& 7 of the table, derived respectively from the 7.46$\mu$m line and from the 12.37$\mu$m line. A small correction has been made for 
extinction at these wavelengths using the coefficients listed in Chiar \& 
Tielens\,\cite{chiar}. Because the extinction coefficients are small at these 
wavelengths, the corrections are insensitive to the precise value. For example,
an extinction \ebv2 at 12.37$\mu$m yields a correction of 10\%. Thus any 
uncertainty in the mid infrared extinction is unimportant. Comparing the cols.6
\& 7, it can be seen that very similar values of H$\beta$ are usually found
from each of the lines. The combined value, given in col.8, is heavily weighted
in favor of the value of H$\beta$ found from the better measured 12.37$\mu$m 
line. The observed value of H$\beta$ is listed in col.9 and is usually taken 
from Acker et al.\cite{acker}. The difference between cols.9 \& 8 is the value 
C$_{ir}$ given in the last column.         

The 16 PNe listed in Table 1 were selected as galactic bulge objects using
three criteria. First and perhaps most important, is the high radial velocity 
of these PNe. Only png356.5-2.3 has a low velocity. This value is shown in the
second column of the table. Secondly the diameter of the nebulae (listed in 
Table 2) is always less than 7\arcsec. Thirdly, to insure that no nearby PNe 
are included, the 6cm radio continuum flux 
density (listed in Table 3) is always less than 70 mJy. In addition it was 
important that additional information concerning the visible spectrum and 
continuum radio flux density be available. It may be noticed that the intrinsic
H$\beta$ flux of these nebulae (column 8 of Table1) is always very similar; in 
all cases it is within a factor of 2 of a value -log H$\beta$=11. The mass of
each PN, computed for the case of a uniform constant density nebula, is also 
shown in Table 2. They are consistent with the nebulae being located in the
Galactic bulge. Nine of these nebulae have been studied by Gutenkunst et 
al.\cite{guten}. We did not use the two further PNe studied by this group 
because no visual spectra are available. Four PNe are from a program to
observe low excitation nebulae. The remaining three PNe (png2.4-3.7, 2.6-3.4 
and 8.2+6.8) were inspired by the work of Hultzsch et al.\,\cite{hultz}. The 
other two bulge PNe studied by these latter authors were not measured by the
{\em Spitzer} Space Telescope.

\begin{table*}[htbp]
\caption[]{H$\beta$ values predicted by infrared hydrogen line and consequent 
values of C$_{ir}$.}

\begin{center}
\begin{tabular}{|l|r|cc|c|cc|c|c|c|}
\hline
\hline
Nebula  & Vel. & I(7.46) & I(12.37) & T$_{e}$ &-log H$\beta$ &-log H$\beta$ & 
predicted & measured &     \\
PNG & km/sec & 10$^{-14}$ & 10$^{-14}$ & K & H(6-5) & H(7-6) &-log H$\beta$ &
-log H$\beta$ & C$_{ir}$ \\
\hline

000.0-6.8 &-84 & 18.8  & 11.9  & 6.000 & 11.29:& 11.01 & 11.05 & 11.99 & 0.94 \\
000.7+3.2 & -175 & 13.5 & 5.35 & 8.500 & 11.34 & 11.25 & 11.27 &13.42 & 2.15 \\
000.7+4.7 & +40 & 36.2 & 14.2 & 8.000  & 10.89 & 10.80 & 10.82 & 13.94 & 3.12\\
001.2+2.1 & -172 & 22.2 & 10.5 & 9.300 & 11.08 & 10.88 & 10.93 & 13.73 & 2.80\\
001.4+5.3 & +42 & 19.4 & 6.70 & 8.300 & 11.19 & 11.18 & 11.18 & 12.71 & 1.53 \\
002.4-3.7 & -83 & 45: & 15.5 & 7.000  & 10.87 & 10.85 & 10.86 & 11.98 & 1.12 \\
002.6-3.4 & +202 &   & 17.0 & 6.000 &        &10.85 & 10.85 & 12.07 & 1.22  \\
002.8+1.7 & +164 & 31.2 & 11.2 & 7.000 &11.01 & 10.94 & 10.96 & 13.48 & 2.52 \\
006.0-3.6 & +136 & 61.7 & 17.1 & 9.600 & 10.67 & 10.75 & 10.73 & 12.19 & 1.46\\
008.2+6.8 & +21.7& 23.1 & 8.45 & 11.000 & 11.10 & 11.05 & 11.07 & 12.13& 1.06\\
351.2+5.2 & -128& 28.0 & 8.10 & 6.000 & 11.11 & 11.17 & 11.15 & 12.12 & 0.97 \\
354.2+4.3 & -75 & 19.7 & 7.58 & 6.750 & 11.23 & 11.16 & 11.18 & 12.62 & 1.44 \\
356.5-2.3 &-16.5& 128.6 & 48.5 & 6.000 &10.43 & 10.37 & 10.39 & 12.23 &1.84 \\
357.1-4.7 &+76 & 13.9: & 9.5 & 6.200 & 11.40:& 11.09 & 11.13 & 12.52 &1.39 \\
358.9+3.2 & +190& 38.8 & 11.9 & 8.500 & 10.88 & 10.90 & 10.90 & 13.03 & 2.13 \\
359.3-1.8 &-89 & 37.4 & 30.9 & 7.000 & 10.89: & 10.46  & 10.52 & 13.95 & 3.43\\

\hline

\end{tabular}
\end{center}

Units of line intensity are always ergs cm$^{-2}$ s$^{-1}$.

\end{table*}

\subsection{The Balmer decrement and extinction}

Selective extinction can be derived from the Balmer decrement assuming that the
hydrogen lines are formed in an optically thin medium and their intrinsic ratios
are given by the theory of Hummer \& Storey \cite{hummer}. Because measurements
of the weaker Balmer lines are more uncertain for the bulge PNe more weight is
usually given to the H$\alpha$/H$\beta$ ratio; sometimes only this ratio is
used. The value of total extinction at H$\beta$, C$_{bd}$, derived from this 
ratio may be written as:\\

C$_{bd}$= X log [I(H$\alpha)$/2.85I(H$\beta$)]                           (1)\\

where 2.85 is the intrinsic value of the ratio H$\alpha$/H$\beta$ and is 
slightly dependent on the electron temperature. 
I(H$\alpha$)/I(H$\beta$) is the measured value of this ratio, while

X=A$_{4861}$/(A$_{4861}$-A$_{6563}$)

and A is the extinction in magnitudes at the wavelength given. These values 
depend on R$_{V}$, the ratio of total to selective extinction. We use the value
R$_{V}$=3.1 in the remainder of this section. Even having fixed the value of
R$_{V}$ there is still a small uncertainty in the value of X to be used. 
Cardelli et al.\cite{ccm} find X=3.36, Seaton \cite{seaton} derives a value of
X=3.23 from measurements of hydrogen and helium lines in NGC7027 (he gives 3.17
as the expected standard value) and Savage \& Mathis \cite{sm} find X=3.17. We 
assume that these values are all within the uncertainty of the determination 
and use in the following the slightly uncertain average value X=3.17.

In Table 2 the values of C$_{bd}$ found in the literature are given. Columns 4
and 5 are the values found using the H$\alpha$/H$\beta$ ratio listed by Acker et al.\,\cite{acker} and Escudero et al.\,\cite{escud} in conjunction with eq.(1).
Columns (6) through (10) give the value of C$_{bd}$ that the authors listed have 
derived from their spectra. Most authors do not give their measured spectral 
intensities but only the intensities after correction for extinction. Thus the
precise extinction correction each individual author used cannot be found. This
adds to the uncertainty. Comparing the different values of C$_{bd}$ for a single
object, we can obtain a feeling as to the accuracy of the individual values, 
which sometimes agree well with one another and sometimes show important 
differences. In the last column the average value of the measures for each PN 
are given.

\begin{table*}[htbp]
\caption[]{ Values of C$_{bd}$ derived from the Balmer decrement.}

\begin{center}
\begin{tabular}{|l|r|c|c|ccccccc|}
\hline
\hline
Nebula & Nebula &Diam.&Mass &C$_{bd}$ &C$_{bd}$ &C$_{bd}$ &C$_{bd}$ & C$_{bd}$&C$_{bd}$ &C$_{bd}$ \\
 PNG &  Name &\arcsec&10$^{-2}$M\smallsun&  (A)$^{\dagger}$ & (B)$^{\dagger}$ & (C)$^{\dagger}$ & (D)$^{\dagger}$ & (E)$^{\dagger}$ &  (F)$^{\dagger}$ & average \\

\hline
000.0-6.8&He2-367& 1.5&3.4&0.95 &0.79 &     &     &     &0.64& 0.85 \\
000.7+3.2&He2-250& 5.6 &22.& 2.19&    & 2.11&     & 2.34&    & 2.20 \\
000.7+4.7& H2-11 & 1.5 &4.8& 3.38&2.90 &    &     &     &    & 3.14 \\
001.2+2.1&He2-262& 2.32&9.5& 2.72 & 2.40&    &     &    &    & 2.68 \\
001.4+5.3& H1-15 & 4.4 &17. &1.27 &    & 1.45&     &    &    & 1.43 \\
002.4-3.7& M1-38 & 3.5 &16. &1.25 &    &     &1.08 &1.17&0.85 & 1.10 \\
002.6-3.4& M1-37 & 4.33&22. & 1.22 &   & 1.15&     &1.23&1.14  & 1.15 \\
002.8+1.7& H2-20 & 3.89 &17.& 3.15 &    &    &     & 2.99&    & 3.11 \\
006.0-3.6& M2-31 & 4.0  &32. & 1.36 &    & 1.41&1.37 &1.42&    & 1.38 \\
008.2+6.8&He2-260& 1.93 &6.1 & 0.99 &0.69&     &     &    &    & 0.82 \\
351.2+5.2& M2-5  & 5.0  &21. & 1.17 &     &     &    & 0.98 &   & 1.11 \\
354.2+4.3& M2-10 & 4.0  &14. &1.81 &      &     &1.32 &1.77 &    & 1.55 \\
356.5-2.3& M1-27 & 6.6  &68. & 2.15&      &1.90 &     &     & 2.23& 2.08 \\
357.1-4.7& H1-43 & 3.   &9.3 & 1.21&0.97 &      &1.07 &    &     & 1.10 \\
358.9+3.2& H1-20 & 4.0  &21. & 2.28&     & 2.29 &     &2.58 & 2.64&2.35 \\
359.3-1.8& M3-44 & 4.0  &30. & 3.30 &      &    &     &3.30 &3.86 &3.42 \\

\hline

\end{tabular}
\end{center}

$^{\dagger}$ References; (A) Acker et al.\,\cite{acker}, (B) Escudero et al.\,\cite{escud}, (C) Cuisinier et al.\,\cite{cuisi}, Wang \& Liu \cite{wang}, (D) Gorny et al.\,\cite{gorny}.\cite{gorny2}, (E) Ratag \cite{ratag}, (F) Exter et al.\,\cite{exter} \\  
$^{\ast}$ C is the extinction used by the author
\end{table*}

\subsection{Comparison of C$_{bd}$ and C$_{ir}$}

Figure 1 is a plot of C$_{bd}$, taken from Table 2, against C$_{ir}$, taken from
Table 1. Approximate error bars are shown although the errors are rather 
difficult to specify accurately. The error in C$_{bd}$ is determined by the 
consistency of the various values in individual objects where several 
measurements have been made. When only one or two measurements are available it
is assumed that the uncertainty is similar to that of the better observed 
object. The uncertainty in C$_{ir}$, which has been discussed above, is
considerably smaller because the 12.37$\mu$m line can be well measured. In 
addition, the uncertainty introduced by a diaphragm correction is probably not
large ($\leq$10\%). On the whole one may say that there is good agreement 
between 
C$_{bd}$, found using a value of $R_{V}$=3.1, and C$_{ir}$ which does not depend 
on $R_{V}$. Thus this value of $R_{V}$ is consistent for these galactic bulge 
nebulae. This consistency removes the basis for the suggestion of Stasinska et 
al.\,\cite{stasinska} that the value of the ratio of the total-to-selective 
extinction $R_{V}$ differs in the material closer to the galactic center from 
the material farther out.

\begin{figure}
  \centering
  \includegraphics[width=7.0cm,angle=90]{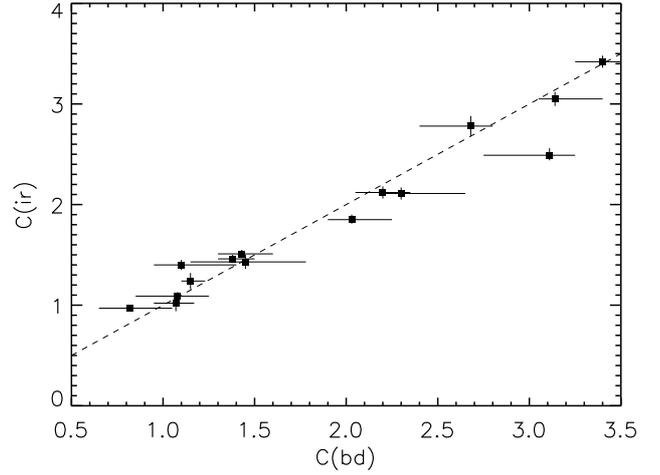}
  \caption{Total extinction coeficient at H$\beta$ derived from the far
infrared hydrogen lines, C$_{ir}$, is plotted as a function of the selective 
extinction coefficient at H$\beta$ derived from the Balmer decrement, C$_{bd}$. 
The dashed line shows when these two quantities are equal. The error bars are
discussed in the text.}
  \label{fig-1}
\end{figure}

\subsection{The 6\,cm radio emission and C$_{rad}$}

Since the flux of the mid-infrared hydrogen lines were not yet known at the 
time of the work of Stasinska et al.\,\cite{stasinska}, these authors used the
PNe radio continuum flux density to determine the intrinsic H$\beta$ flux. As
discussed earlier, in order to compute this intrinsic flux, one must know the 
electron temperature T$_e$ as well as the amount of He$^+$ and He$^{++}$, which
are also responsible for the continuous radio emission. Furthermore the nebula
must be optically thin to the radio radiation. The electron temperature (taken
from an as yet unpublished study of Pottasch \& Bernard-Salas) is 
listed in Table 1 as well as well as the ionized helium abundances in the cols. 2 and 3 of Table 3. The 6\,cm flux density is listed in col. 4 of Table 4; it 
has been measured mainly by Gathier et al.\cite{gathier}, and also by Zijlstra 
et al.\,\cite{zijlstra}, Pottasch et al.\,\cite{pott1} and Aaquist \& Kwok\,\cite{aaquist}. The only nebula with an uncertain radio flux density is H2-20 which 
was only measured at 3\,cm by Purton et al.\,\cite{purton}. It has a large 
error. The 6\,cm flux density of this PN is discussed by Cahn et al.\,\cite{cahn}, whose (uncertain) value is listed in Table 3. He2-367 is not on the list 
since no measurements of its radio continuum flus density have been made.

\begin{table}[htbp]
\caption[]{ Values of C$_{rad}$ derived from the 6\,cm radio continuum.}

\begin{center}
\begin{tabular}{|l|cccccc|}
\hline
\hline
Nebula &He$^+$/H &He$^{++}$/H &S$_{6cm}$ &S$_{21cm}$&-log H$\beta$ &C$_{rad}$ \\
 PNG   &         &         &   mJy &   mJy &    &   \\
\hline
000.7+3.2 &  0.12  & 0.032 &   15. &  15.6  & 11.31 &  2.11 \\
000.7+4.7 &  0.178 &  0    &  27.7 &  11.5  & 11.00 & 2.94 \\
001.2+2.1 &  0.11  & 0.001 &  26.  &  24.1  & 11.07 & 2.66 \\
001.4+5.3 &  0.103 &  0    &  13.  &  13.6  & 11.34 & 1.37 \\
002.4-3.7 &  0.0125 &  0   &  24.  &  14.9  & 10.97 & 1.04 \\
002.6-3.4 &  0.0085 &  0   &  15.  &  10.9  & 11.19 & 0.88 \\
002.8+1.7 &  0.075  &  0   &  16.3 &  13.1  & 11.21 & 2.27 \\
006.0-3.6 &  0.12  &   0   &  51.  &  41.2  & 10.78 & 1.41 \\
008.2+6.8 &  0.009 &   0   &  13.  &   8.1  & 11.39 & 0.74 \\
351.2+5.2 &  0.076 &   0   &  12.  &  14.1  & 11.30 & 0.82 \\
354.2+4.3 &  0.135 &   0   &   9.1 &  11.6  & 11.45 & 1.17 \\
356.5-2.3 & 0.020  &   0   &  63.  &  65.4  & 10.64 & 1.59 \\
357.1-4.7 & 0.020  &   0   &   6.3 &   4.4  & 11.56 & 0.96 \\
358.9+3.2 &  0.126 &   0   &  32.  &  27.3  & 10.99 & 2.04 \\
359.3-1.8 & 0.1    &   0   &  35.  &  25.3  & 10.88 & 3.02 \\

\hline 

\end{tabular}
\end{center}

Helium ionic abundances are from Pottasch \& Bernard-Salas (in prep.).
\end{table}

 The resultant H$\beta$ flux is listed in col.6 of Table 3 and the resultant 
H$\beta$ extinction, C$_{rad}$ in col.7. These values of extinction have been 
plotted against the Balmer decrement extinction C$_{bd}$ as Fig.2.  Comparing
this figure with Fig.1 shows that the points are on average somewhat to the
right of the line of equal extinctions, i.e. the extinction derived from the
radio flux density is somewhat lower than that indicated by the Balmer 
decrement. It is unlikely that this can be due to the use of an incorrect value
of R$_V$ because it has already been established in the previous section that 
R$_V$ must be close to a value of 3.1. The problem must lie in the value of 
the 6\,cm radio flux 
density. While it is possible that these measurements are incorrect, this is 
not likely because the errors would all have to be in the direction as to make
the flux smaller than the actual value. It is more likely that the assumption 
that the nebulae are optically thin is not entirely correct because some
absorption would make the fluxes of all PNe smaller. The following 
evidence supports this assumption. When the nebula is optically thin up to a 
wavelength of 21cm, the 21cm continuum flux density will be approximately a 
factor of 1.14 higher than the 6cm flux density. To check this, col.5 of Table 3
lists the 21cm flux density of these PNe, taken from the VLA measurements of 
Condon \& Kaplan\,\cite{condon}. As can be seen, in only 5 of the 15 nebulae is
the radio emission higher at 21cm than at 6\,cm. Thus optical depth effects are 
playing a role in this radio spectrum. Not enough information is available to
quantify this effect for the individual nebulae, because very little is known
about the distribution of matter in these PNe. The effect will always work 
in the direction of decreasing the predicted H$\beta$ intensity and thus 
increasing C$_{rad}$. This can be settled by obtaining radio measurements at a 
higher frequency.

In Fig.2 the Galactic bulge PNe used by Ruffle et al.\,\cite{ruffle} have also 
been plotted (open circles). Unfortunately less information is known for these 
nebulae (the electron temperature and the helium ionic abundances are unknown) 
and the only information about the Balmer decrement is that in the catalog of 
Acker et al.\,\cite{acker}, so that the average values used by Ruffle et 
al.\,\cite{ruffle} were used and the extinction was obtained from the Balmer 
decrement as in Sect.2.2. There is thus a somewhat greater uncertainty in these
values. However, as can be seen from the figure there is no difference between 
the results from our PNe and those of Ruffle et al.\,\cite{ruffle}.

\begin{figure}
  \centering
  \includegraphics[width=7.0cm,angle=90]{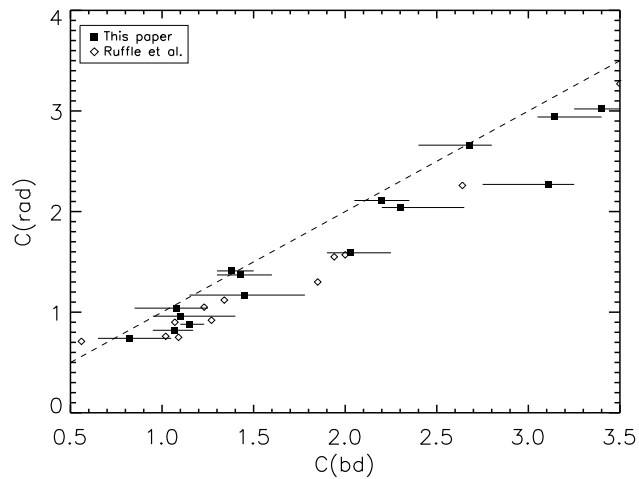}
   \caption{Total extinction coefficient at H$\beta$ derived from the 6cm
radio continuum flux density, C$_{rad}$, is plotted as a function of the 
selective extinction coefficient at H$\beta$ derived from the Balmer decrement, C$_{bd}$. The dashed line shows when these two quantities are equal. In addition
to the PNe discussed in this paper, those discussed by Ruffle et al.\,\cite{ruffle} are also plotted to demonstrate that these PNe lead to the same result. }
  \label{fig-2}
\end{figure}

\section{Conclusions}

The H$\beta$ flux has been derived from the mid-infrared hydrogen lines observed
by the {\em Spitzer} Space Telescope for 16 planetary nebulae in the Galactic 
bulge. These transitions, H(7-6) at 12.37$\mu$m and H(6-5) at 7.45$\mu$m, being
relatively unaffected by galactic extinction, make it possible to obtain the 
total extinction at H$\beta$. We emphasize that this is the most reliable 
method of obtaining the extinction and therefore  C$_{ir}$  is more reliable 
than C$_{rad}$. By combining C$_{ir}$ with the selective extinction 
obtained from measurements of the Balmer decrement for these same PNe, the 
value of the total-to-selective extinction, R$_V$, can be approximately
determined. The value found is consistent with R$_V$=3.1 which is the value of
total-to-selective extinction ususally found in the interstellar material
closer to the Sun.

This result is at odds with that of Stasinska et al.\,\cite{stasinska} and of
Ruffle et al.\,\cite{ruffle}. These authors obtained the total extinction by
using the 6\,cm radio continuum to determine the extinction-corrected H$\beta$
flux. We discuss the reasons for this difference, which most likely is largly
due to the assumption that the PNe are entirely optically thin at this 6\,cm
radio frequency.

\begin{acknowledgements}

We acknowledge the use of SIMBAD and ADS in this research work. J.B-S wishes to
acknowledge the support from a Marie Curie Intra-European Fellowship within the
7th European Community Framework Program under project number 272820.

\end{acknowledgements}

\end{document}